\documentclass{ws-procs9x6}

\begin{document}

\title{PROGRESS IN CORRELATION FEMTOSCOPY\footnote{
\uppercase{W}ork
supported by grant 202/01/0779 of the \uppercase{G}rant
\uppercase{A}gency of the \uppercase{C}zech \uppercase{R}epublic.}}

\author{R.~LEDNICKY}

\address{Joint Institute for Nuclear Research\\
Dubna, Moscow Region, 141980, Russia\\
Institute of Physics ASCR \\
Na Slovance 2,
10800 Prague, Czech Republic\\
E-mail: lednicky@fzu.cz}


\maketitle

\abstracts{
Recent results on particle momentum and spin correlations
are discussed, particularly, in view of the role played by
the effect of final state interaction.
It is demonstrated that this effect allows for
(i) correlation femtoscopy with unlike particles;
(ii) study of the relative space--time asymmetries in the
production of different particle species (including relative
time delays);
(iii) study of the particle strong interaction hardly
accessible by other means (e.g., in $\Lambda\Lambda$ system).
}

\section{Introduction}

The momentum correlations of particles at small relative velocities
are widely used to study space-time characteristics of the
production processes, so serving as a correlation femtoscope.
Particularly, for non-interacting identical particles, like photons or,
to some extent, pions, these correlations result from the
interference of the two--particle amplitudes due to the
symmetrization requirement of quantum statistics
(QS).\cite{GGLP60,KP72}\footnote{
There exists\cite{KP72,KP75} a deep analogy of the momentum QS
correlations of photons with the space--time correlations
of the intensities of classical electromagnetic fields used in
astronomy to measure the angular radii of stellar objects
based on the superposition principle (HBT effect).\cite{hbt}
This analogy is sometimes misunderstood and the momentum
correlations are mixed up with the HBT correlations
in spite of their orthogonal character and the 
absence of the classical analogy
for correlations of identical fermions.
}
The momentum QS correlations were first
observed as an enhanced production of the
pairs of identical pions with small opening angles
(GGLP effect\cite{GGLP60}). Later on,
Kopylov and Podgoretsky\cite{KP72}
settled the basics of correlation femtoscopy; particularly, they
suggested to study the interference effect in terms of the correlation
function and clarified the role of the space--time characteristics
of particle production in various physical situations.

The momentum correlations of particles emitted at nuclear distances
are also influenced by the effect of
final state interaction (FSI).\cite{koo,ll1}
Though the FSI effect complicates the correlation analysis,
it is an important source of information allowing for the
coalescence femtoscopy (see, {\it e.g.},\cite{lyu88} and the
talk by G.~Melkumov),
the correlation femtoscopy with unlike particles\cite{ll1,BS86}
including the access to the relative space--time
asymmetries in particle production\cite{LLEN95} and a study
of particle strong interaction.

The two-particle
correlation function ${\mathcal R}(p_{1},p_{2})$
is usually defined as a
ratio of the measured two-particle distribution to the
reference one obtained by mixing the particles from different
events.
It can be calculated\cite{koo}$^-$\cite{BS86} 
as a square of the properly symmetrized 
stationary solution $\psi ^{S(+)}_{-{\bf k}^*}({\bf r}^*)$ of the
scattering problem averaged over the relative distance ${\bf r}^*$ of
the emitters in the pair c.m.s. 
(${\bf k}^*={\bf p}_1^*=-{\bf p}_2^*\equiv {\bf Q}/2$)
and over the pair total spin $S$.

It is well known that the directional and velocity
dependence of the correlation function
can be used to determine both
the duration of the emission process and the form
of the emission region,\cite{KP72} as well as - to reveal
the details of the
production dynamics (such as collective flows;
see, {\it e.g.},\cite{PRA84}).
The recent (puzzling) results on like pion correlations from
BNL RHIC have been
presented at this conference by M.~Lisa and V.~Okorokov.

\section{Femtometry with unlike particles}
The complicated dynamics of particle production,
including resonance decays and particle rescatterings,
leads to essentially non--Gaussian tail of the
$r^*$--distribution.
Therefore, due to different $r^*$--sensitivity
of the QS, strong and Coulomb FSI effects,
one has to be careful when analyzing the correlation functions
in terms of simple models.
Thus, the QS and strong FSI effects are influenced by the $r^*$--tail
mainly through the suppression parameter $\lambda$ while,
the Coulomb FSI is sensitive to the distances as large as
the pair Bohr radius $|a|$ (hundreds fm for the pairs
containing pions).
These problems can be at least partially overcome with the help of
transport code simulations accounting for the dynamical evolution
of the emission process and providing the phase space information
required to calculate the QS and FSI effects on the correlation function.

Thus, in a preliminary analysis of the NA49 correlation data from
central $Pb+Pb$ 158 AGeV collisions,\cite{lna49}
the transport RQMD v.2.3 code was used. 
To account for a possible
mismatch in $\langle r^*\rangle$, the correlation functions
were calculated with the space--time coordinates 
of the emission points scaled by 0.7, 0.8 and 1. 
The scale parameter was then fitted
using the quadratic interpolation.
The fits of the $\pi^+\pi^-$, $\pi^+ p$ and $\pi^- p$ 
correlation function indicate that RQMD
overestimates the distances $r^*$ by 10-20$\%$. 

Recently, there appeared data on $p\Lambda$ correlation functions
from $Au+Au$ experiment E985 at AGS.\cite{lis01}
As the Coulomb FSI is absent
in this system, one avoids here the problem of its sensitivity
to the $r^*$--tail. Also, the absence of the Coulomb suppression
of small relative momenta makes this system more sensitive to the
radius parameters as compared with $pp$ correlations.\cite{wan99}
In spite of rather large statistical errors,
a significant enhancement is seen at low relative momentum,
consistent with the known singlet and triplet
$p\Lambda$ s--wave scattering lengths.
In fact, using the analytical expression for the correlation
function (originally derived for $pn$ system\cite{ll1}),
one gets a good fit of
the combined (4, 6 and 8 AGeV) correlation function
with the Gaussian radius $r_0= 4.5 \pm 0.7$ fm,\cite{lna49}
in agreement with the radii
of 3-4 fm obtained from $pp$ correlations in
heavy ion collisions at GSI, AGS and SPS energies.

\section{Accessing particle strong interaction}
In case of a poor knowledge of the two--particle strong interaction,
which is the case for {\it exotic} systems like ($M=$ meson)
$MM$, $M \Lambda$ or $\Lambda \Lambda$,
the correlation measurements can be also used to study the latter.

In heavy ion collisions, the effective radius $r_0$ of the emission region
can be considered much larger than the range of the
strong interaction potential.
The FSI contribution to the correlation function
is then independent of the actual
potential form.\cite{ll1,gkll86} At small $Q=2k^*$,
it is determined by the s-wave
scattering amplitudes $f^S(k^*)$.\cite{ll1}
In case of $|f^S|>r_0$, this contribution is of the order of
$|f^S/r_0|^2$ and dominates over the effect of QS. In the opposite case,
the sensitivity of the correlation function to the scattering
amplitude is determined by the linear term $f^S/r_0$.

The possibility of the correlation measurement
of the scattering amplitudes has been demonstrated\cite{lna49}
in a recent analysis of the
NA49 $\pi^+\pi^-$ correlation data within the RQMD model.
The fitted strong interaction scale,
redefining the original scattering length $f_0=$ 0.232 fm,
appeared to be significantly
lower than unity: $0.63\pm 0.08$.
To a similar shift ($\sim 20\%$)
point also the recent BNL data on $K_{l4}$ decays.\cite{pis01}
These results are in agreement with the two--loop calculation
in the chiral perturbation theory with a standard value of the
quark condensate.\cite{col00}

As for the $\Lambda\Lambda$ system, the singlet
$\Lambda\Lambda$ s--wave scattering length $f_0$
has been recently estimated\cite{lna49,blu02} based on
the NA49 data on
$\Lambda\Lambda$ correlations in $Pb+Pb$ collisions at
158 AGeV.
Using the analytical expression for the correlation
function and fixing the purity of direct
$\Lambda$--pairs at the estimated value of 0.16 and varying
the effective radius $r_0$ in the acceptable range of several fm,
one gets\cite{blu02}
{\it e.g.}, $f_0=2.4\pm 2.1$ and $3.2\pm 5.7$ fm 
for $r_0=2$ and 4 fm respectively
(we use the same sign convention as for
meson--meson and meson--baryon systems).
Though the fit results are not very restrictive, they
likely exclude
the possibility of a large positive singlet scattering length
comparable to that of $\sim$20 fm for the two--nucleon system.

The important information is
also coming from $\Lambda\Lambda$ correlations at LEP.\cite{ale00}
Here the effective radius $r_0$ is substantially smaller than the
range of the strong interaction potential, so the $\Lambda\Lambda$
correlation function is sensitive to the potential form.
In fact, the observed strong decrease of the correlation function at small
$Q$ can be considered as a direct evidence for the potential core;\cite{led02}
particularly, the Nijmegen potential NSC97e yields a reasonable agreement
with this data.

\section{Accessing relative space-time asymmetries}

The correlation function of two non--identical particles,
compared with the identical ones,
contains a principally new piece of information on the relative
space-time asymmetries in particle emission.\cite{LLEN95}
Since this information enters in the two-particle amplitude
$\psi_{-{\bf k}^{*}}^{S(+)}({\bf r}^{*})$
through the terms odd in
${\bf k}^*{\bf r}^*\equiv {\bf p}_1^*({\bf r}_1^*-{\bf r}_2^*)$,
it can be accessed studying the correlation functions
${\mathcal R}_{+i}$ and ${\mathcal R}_{-i}$
with positive and negative projection $k^*_i$ on
a given direction ${\bf i}$ or, - the
ratio ${\mathcal R}_{+i}/{\mathcal R}_{-i}$.
For example, ${\bf i}$ can be the direction of the pair
velocity or, any of the out (x), side (y), longitudinal (z)
directions. Note that in the longitudinally comoving system (LCMS),
one has $r^*_i=r_i$ except for
$r_x^*\equiv\Delta x^*=\gamma_{t}(\Delta x-v_{t}\Delta t)$,
where $\gamma_{t}$ and $v_t$
are the pair LCMS Lorentz factor and velocity.
One may see that the asymmetry in the out (x) direction
depends on both space and time asymmetries
$\langle\Delta x\rangle$ and $\langle\Delta t\rangle$.
In case of a dominant Coulomb FSI, the intercept of the correlation
function ratio is directly related with the asymmetry
$\langle r^*_i\rangle$ scaled by the Bohr radius
$a=(\mu z_{1}z_{2}e^{2})^{-1}$:
$
{\mathcal R}_{+i}/{\mathcal R}_{-i}\approx 1+
2\langle r_i^*\rangle /a.
$

A review of the simulation studies of the method sensitivity and
the experimental results can be found elsewhere\cite{lna49}.
Here we discuss the out correlation asymmetries observed
for $\pi p$ and $\pi K$ systems in heavy ion collisions at CERN
SPS and BNL RHIC.\cite{lna49,ret01}
These asymmetries are in agreement with practically
charge independent meson production and a negative
$\langle\Delta x\rangle$
or positive $c\langle\Delta t\rangle$ on the level of
several fm (assuming $m_1<m_2$).
The RHIC asymmetries seem to be overestimated by the
RQMD model while
the NA49 $\pi^+\pi^-$ and $\pi p$ asymmetries
in central $Pb+Pb$ collisions
at 158 AGeV are in quantitative agreement with this model -
it yields practically zero asymmetries for $\pi^+\pi^-$ system
while, for $\pi^\pm p$ systems,
$\langle\Delta x\rangle \doteq -5.2$ fm,
$\langle\Delta t\rangle \doteq 2.9$ fm/c,
$\langle\Delta x^*\rangle \doteq -8.5 fm$.
Besides, it predicts $\langle x\rangle$ increasing 
with particle $p_t$ or
$u_t=p_t/m$, starting from zero due to kinematic reasons.
The asymmetry arises because of a faster increase with $u_t$
for heavier particle.
In fact, the hierarchy
$\langle x_\pi\rangle<\langle x_K\rangle<\langle x_p\rangle$
is a signal of a universal transversal collective
flow;\cite{lna49}
one should simply take into account that the mean thermal
velocity is smaller for heavier particle
and thus washes out the positive shift due to the flow to a lesser extent.

\section{Spin correlations}

The information on the system size and the two--particle interaction
can be achieved also with the help of spin correlation measurements
using as a spin analyzer the asymmetric (weak) particle
decay.\cite{ale95}$^-$\cite{ll01}
Since this technique requires no construction of the uncorrelated
reference sample, it can serve as an important consistency check
of the standard correlation measurements.
Particularly, for two $\Lambda$--particles decaying into the $p\pi^-$
channel,
the distribution of the cosine of the relative angle $\theta$ between
the directions of the decay protons in the respective $\Lambda$ rest
frames allows one to determine the triplet fraction
$\rho_t={\mathcal R}_t/{\mathcal R}$, where ${\mathcal R}_t$ is
the triplet part of the correlation function.

The spin correlations allow also for a relatively simple
test of the quantum--mechanical coherence based on Bell--type
inequalities derived from the assumption of the factorizability
of the two--particle density matrix, i.e. its reduction to a sum
of the direct products of one--particle density matrices with the
nonnegative coefficients.\cite{ll01} Clearly, such a form of the
density matrix corresponds to a classical probabilistic description and
cannot account for the coherent quantum--mechanical effects,
particularly, for the production of two $\Lambda$-particles in a
singlet state. Thus
the suppression of the triplet $\Lambda\Lambda$ fraction observed
in multihadronic $Z^0$ decays at LEP\cite{ale00}
indicates a violation of one of the Bell-type inequalities,
$\rho_t \ge 1/2$.

\section{Conclusions}

The particle momentum and, recently, also spin correlations 
give unique information on the space--time
production characteristics including collective flows.
Rather direct evidence for a strong transverse flow in heavy ion
collisions at SPS and RHIC is coming from unlike particle correlation
asymmetries. Being sensitive to relative time delays and collective flows,
the correlation asymmetries can be especially useful to study
the effects of the quark--gluon plasma phase transition.
The correlations yield also a valuable information on the
particle strong interaction hardly accessible by other means.

\end{document}